\documentclass{article}

\usepackage{mathptmx}
\usepackage{amsmath}
\usepackage{graphicx}

\usepackage[utf8]{inputenc} 

\begin{document}

\title{Influence of parallel computing strategies of iterative imputation of
missing data: a case study on missForest}


\author{Shangzhi Hong, Yuqi Sun, Hanying Li, Henry S. Lynn}


\begin{titlepage}
  \maketitle
\end{titlepage}

\begin{abstract}
  Machine learning iterative imputation methods have been well accepted by
  researchers for imputing missing data, but they can be time-consuming when
  handling large datasets. To overcome this drawback, parallel computing
  strategies have been proposed but their impact on imputation results and
  subsequent statistical analyses are relatively unknown. This study examines
  the two parallel strategies (variable-wise distributed computation and
  model-wise distributed computation) implemented in the random-forest
  imputation method, missForest. Results from the simulation experiments
  showed that the two parallel strategies can influence both the imputation
  process and the final imputation results differently. Specifically,
  even though both strategies produced similar normalized root mean squared
  prediction errors, the variable-wise distributed strategy led to additional
  biases when estimating the mean and inter-correlation of the covariates and
  their regression coefficients.
\end{abstract}

%
%
%
%
\section{Introduction}
\label{sec:1}
Missing data are common in most research, and various kinds of imputation
methods have been proposed for handling missing data problems. Stekhoven and
Buhlmann (Stekhoven and Buhlmann 2012) proposed the missForest algorithm based
on a random forest (RF) machine learning method (Liaw and Wiener 2002), and it
has been used in different studies and benchmarked against other imputation
methods (Ramosaj and Pauly 2019; Shah et al. 2014; Tang and Ishwaran 2017;
Waljee et al. 2013).

Moreover, new imputation methods have been proposed based on the original
missForest algorithm (Mayer 2019). MissForest has been shown to have superior
predictive accuracy under certain circumstances, but it necessitates the
building of a large number (default is 100) of trees during the imputation
process for a single variable per iteration and usually several iterations
are required. Likewise, missForest can be computationally intensive and
time-consuming for large datasets, thereby limiting its usability.
To boost performance, two parallel computing strategies (referred to as
“forests” and “variables” in the software package) suitable for “long” and
“wide” datasets, respectively (Stekhoven 2013), were implemented in missForest
 with the release of version 1.4. However, there has not been any published
 evaluation of their difference on predictive accuracy and subsequent impact
 on statistical analyses. The implicit assumption is that these two strategies
 are equally valid and will lead to similar results.

This study uses simulation experiments to address the differences between
these two parallel computation strategies of missForest. Computational
efficiency can be critical for handling large datasets; thus, this study’s
results can be of use to both data analytics practitioners and methodologists
for imputation methods.
%
%
%
%
\section{Methods}
\label{sec:2}
\subsection{MissForest algorithm}
\label{sec:2.1}
In missForest, the variables containing missing values are initialized for
imputation by replacing the missing cells by corresponding mean values (for
continuous variables), or by the most frequent category (for categorical
variables). A variable under imputation is then divided into two distinct
parts: the observed part that contains no missing values, and the missing part
that serves as the prediction set. A random forest is fitted using the observed
part as response and the corresponding values of the other variables as
predictors, and the missing part is replaced with the predicted values from the
random forest. The algorithm then proceeds to the next variable to be imputed,
and the iteration stops when the difference between the current and previously
imputed values increase or if the maximum number iteration is reached. Since
the release of missForest version 1.4, two parallel strategies have been
implemented to increase the computational efficiency when applying random forest
imputation to large datasets.

\subsubsection{Strategy 1: distributing the computation of imputing a single
variable}
\label{sec:2.1.1}
In the first strategy, the building of the ensemble of trees for a variable to
be imputed are divided into smaller subsets and distributed into different
computing processes based on the number of core processors in the computer.
The results from different ensembles of smaller trees are recombined into a
single one, and the final predictions are derived from the combined ensemble of
trees. Each variable to be imputed undergoes this process until all the
variables have been imputed in a single iteration. This strategy is most useful
if the process of building a random forest is time-consuming and the number of
variables in the dataset is relatively small.

\subsubsection{Strategy 2: distributing the computation of different variables}
\label{sec:2.1.2}
In the second strategy, the computation of the random forest for each variable
to be imputed in a single iteration is distributed to different computing
processes. The imputations of the variables are done simultaneously and
independent of each other with the building of the ensemble of trees for each
variable performed by a single process. After all the variables have been
imputed, the results are recombined to form a single complete dataset.
The current iteration is then finished, and the algorithm moves to the next
iteration. This strategy can be useful for datasets containing many variables
while the time consumption for building the random forest for a single variable
is small.

\subsection{Simulation studies}
\label{sec:2.2}
To further investigate the influence of the choice of parallel strategies on
imputation, a series of simulations and analyses were carried out using R,
version 3.6 (R Core Team, Vienna, Austria) (R Core Team 2019).
Four sequential stages were involved:
\begin{enumerate}
  \item Data generation: complete datasets were simulated based on pre-defined
scenarios.
  \item Amputation: the complete datasets were made incomplete based on
specified rules.
  \item Imputation: the missing values contained in the simulated datasets were
filled in by missForest using different parallel strategies.
  \item Analysis: Statistical analysis were performed on both the original
complete datasets and the corresponding imputed datasets, and comparisons
were made.
\end{enumerate}

\subsubsection{Data generation}
\label{sec:2.2.1}
The data structures were made as simple as possible with a response Y and just
two covariates $X_{1}$ and $X_{2}$ to enhance the investigation of the influence
of the two parallel strategies on imputation results. Also, a large variance was
used to get more discriminative results. Three different sets of 2000 simulated
datasets containing 200 observations each were generated based on following
settings:
\begin{enumerate}
  \item Uncorrelated covariates with linearly dependent response:
    \begin{equation}
      \left[\begin{array}{l}
      {Y} \\
      {X_{1}} \\
      {X_{2}}
      \end{array}\right]
      \sim \text { Normal }
      \left(\left[\begin{array}{l}
      {2} \\
      {1} \\
      {1}
      \end{array}\right],\left[\begin{array}{ccc}
      {21} & {10} & {10} \\
      {10} & {10} & {0} \\
      {10} & {0} & {10}
      \end{array}\right]\right)
    \end{equation}
    And the conditional distribution (Searle and Gruber 2016) of $Y$ given
    $X_{1}=x_{1}$ and $X_{2}=x_{2}$
    is
    \begin{equation}
      \left(Y | X_{1}=x_{1}, X_{2}=x_{2}\right)
      \sim \text { Normal }
      \left(x_{1}+x_{2}, 1\right)
    \end{equation}
  \item Correlated multivariate normal data:
    \begin{equation}
      \left[
      \begin{array}{l}
      {Y} \\
      {X_{1}} \\
      {X_{2}}
      \end{array}\right]
      \sim \text{Normal}
      \left(\left[\begin{array}{l}
      {1} \\
      {1} \\
      {1}
      \end{array}\right],
      \left[\begin{array}{ccc}
      {10} & {10 \rho} & {10 \rho} \\
      {10 \rho} & {10} & {10 \rho} \\
      {10 \rho} & {10 \rho} & {10}
      \end{array}\right]\right)
    \end{equation}
    The correlation coefficients were $\rho=0.25$ or $\rho=0.75$, roughly
  corresponding to weakly correlated and strongly correlated data.
  This multivariate distribution leads to the following conditional
  distributions of $Y$ given $X_1=x_1$ and $X_2=x_2$:
  \begin{equation}
    \left(Y | X_{1}=x_{1}, X_{2}=x_{2}\right)
    \sim \text { Normal }
    \left(0.2x_{1}+0.2x_{2}+0.6, 9\right)
  \end{equation}
  and
  \begin{equation}
    \left(Y | X_{1}=x_{1}, X_{2}=x_{2}\right)
    \sim \text { Normal }
    \left(\frac{3}{7}x_{1}+\frac{3}{7}x_{2}+\frac{1}{7}, \frac{25}{7}\right)
  \end{equation}
\end{enumerate}
Altogether, the simulation consists of three different data generation
scenarios: (1) multivariate normal data with independent covariates,
(2) moderately correlated multivariate normal data, and
(3) strongly correlated multivariate normal data.

\subsubsection{Amputation}
\label{sec:2.2.2}
Amputation functions (Schouten et al. 2018) provided by the “MICE”
(van Buuren and Groothuis-Oudshoorn 2011) R package were used in this study
to generate missing values. Missing at random (MAR) patterns were introduced
by setting $X_{1}$ and/or $X_{2}$ to be missing depending on $Y$. Specifically,
the probability of each observation being missing was set to 50\% according to
a standard right-tailed logistic function on $Y$; thus the probability of
the covariates being missing is higher for observations with higher values
of $Y$. Two MAR patterns are generated, whereby either both covariates are
missing (i.e., two missing cells) or only one of the covariates is missing
(i.e., one missing cell.)

\subsubsection{Imputation}
\label{sec:2.2.3}
The amputed datasets underwent imputation by missForest, and default parameter
values (number of trees grown was set to 100, and maximum iteration was set
to 10) were accepted as recommended by the original article
(Stekhoven and Buhlmann 2012). The number of distributed computing processes was
set to three, which equals to the number of variables in the dataset
(the maximum allowed by missForest), to allow for more computing resources
available for “forests” strategy. Imputation without parallelization,
parallelized imputation by forests and by variables were performed.

\subsubsection{Analysis}
\label{sec:2.2.4}
Comparisons were made between the two parallel strategies, along with the
original sequential algorithm, based on:
\begin{enumerate}
  \item the number of iterations performed using different parallel strategy;
  \item relative bias for the mean and for the standard deviation of the imputed
  variable:
  \begin{equation}
    \frac{\operatorname{mean}
    \left(\boldsymbol{V}_{\mathrm{imp}}\right)}
    {\operatorname{mean}
    \left(\boldsymbol{V}_{\mathrm{true}}\right)}-1
  \end{equation}
  and
  \begin{equation}
    \frac{\operatorname{sd}
    \left(\boldsymbol{V}_{\mathrm{imp}}\right)}
    {\operatorname{sd}
    \left(\boldsymbol{V}_{\mathrm{true}}\right)}-1
  \end{equation}
  where $\boldsymbol{V}$ is either one of the imputed variables ($X_{1}$ or
  $X_{2}$), $\boldsymbol{V}_{true}$ is the original vector of true values,
  $\boldsymbol{V}_{imp}$  is the data vector after imputation, and the mean and
  standard deviation are computed over all the data values.
  \item the relative bias of the coefficient estimate:
    \begin{equation}
      \frac{\left(\widehat{\beta}_{p}-\beta_{p}\right)}{\beta_{p}},
      p = 1, 2, \text { or } 3
    \end{equation}
    corresponding to the intercept (if any), $X_{1}$ or $X_{2}$.
  \item normalized root mean squared error (NRMSE) value:
  \begin{equation}
    \sqrt{
      \left.
      \frac
      {
          \operatorname{mean}
          \left(\left(\boldsymbol{X}_{\text {true}}-
          \boldsymbol{X}_{\text {imp}}\right)^{2}\right)
      }
      {
      \operatorname{var}\left(\boldsymbol{X}_{\text {true }}\right)}
      \right.
      }
  \end{equation}
  where $\boldsymbol{X}_{\text {true}}$  and $\boldsymbol{X}_{\text {imp}}$ are
  the true and imputed data matrix, respectively, and the mean and variance
  are computed only over the missing values.
\end{enumerate}
Pearson correlation coefficients were also estimated for certain data scenarios
when investigating the influence of imputation on the relationships between
imputed variables. If the two parallel algorithms are equivalent and valid,
then their imputation results should not be dissimilar with the sequential
algorithm in imputation accuracy for all four criteria.
\section{Results}
\label{sec:3}
The results from three different parallel strategies showed variations in
iteration numbers, relative bias of sample mean, relative differences of
standard derivation and regression estimates in linear regression data scenario.
However, such differences showed correlated relationship with different data
scenarios.
\subsection{Iterations performed}
\label{sec:3.1}
\begin{figure}[h!]
  \includegraphics[width=\linewidth]{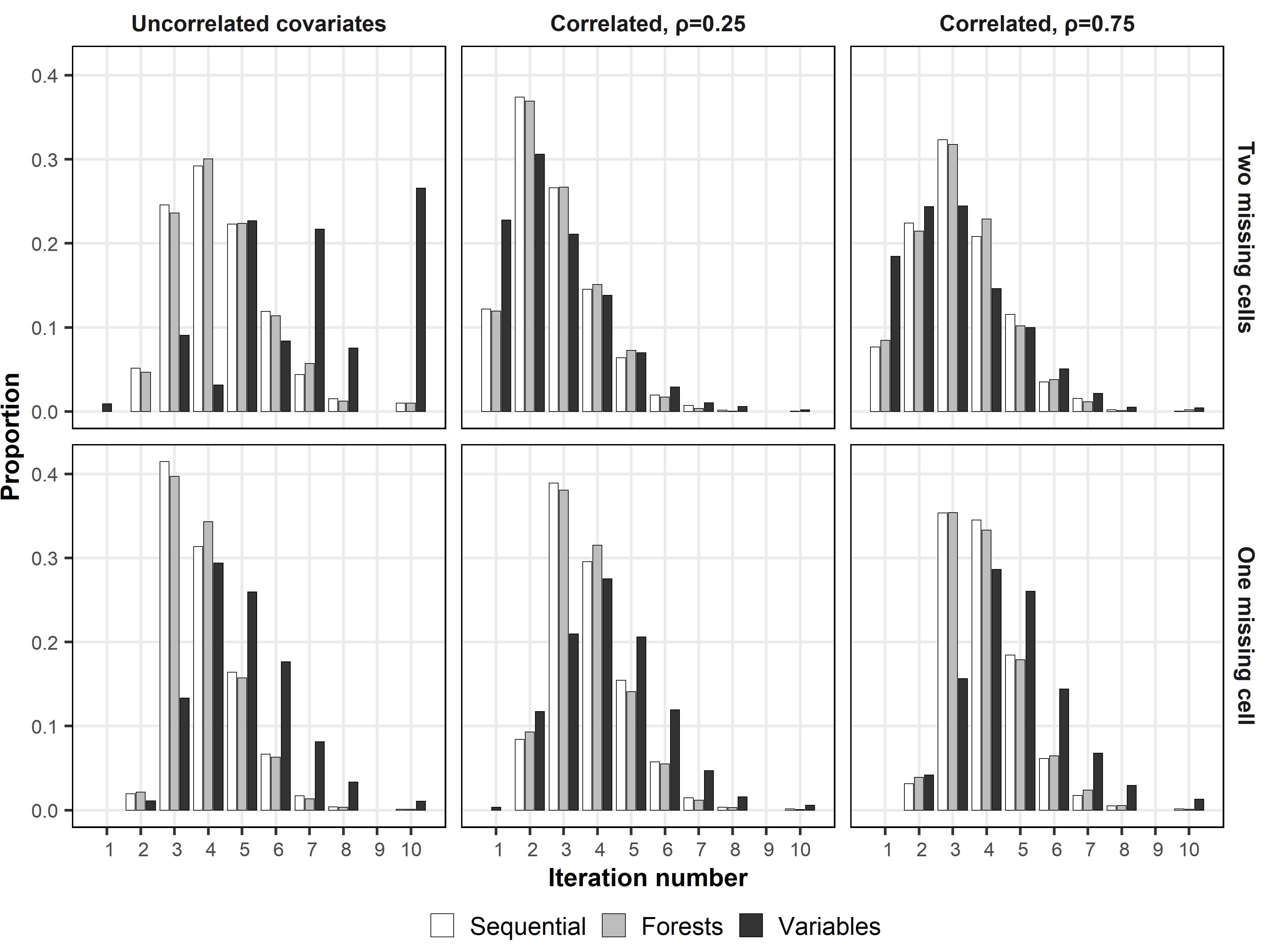}
  \caption{Number of iterations for different parallel modes in different
data generation settings and missing patterns}
  \label{fig:1}
\end{figure}
The number of iterations of imputation with the “variables” parallel strategy
was very different (p\textless0.001 across all scenarios, Fisher’s exact test)
from the other two strategies for all eight data scenarios, while sequential
imputation and parallel “forests” strategies were more similar. For the
parallel “forests” and sequential strategies, most imputation runs stopped
at two to four iterations with only a small number of runs (\textless0.25\%
overall) reaching the maximum number of ten loops. For the “variables” parallel
imputation, however, imputation often require larger number of iterations
even for data with one missing cell per observation (median no. of
iterations = 3, 4 for two missing cells, one missing cell, respectively.)
Note also that with two missing cells per observation, an exceedingly large
proportion of runs (26.5\%) stopped at the maximum iteration number for the
“variables” strategy (Fig. 1).

\subsection{Relative bias of sample mean}
\begin{figure}[h!]
  \includegraphics[width=\linewidth]{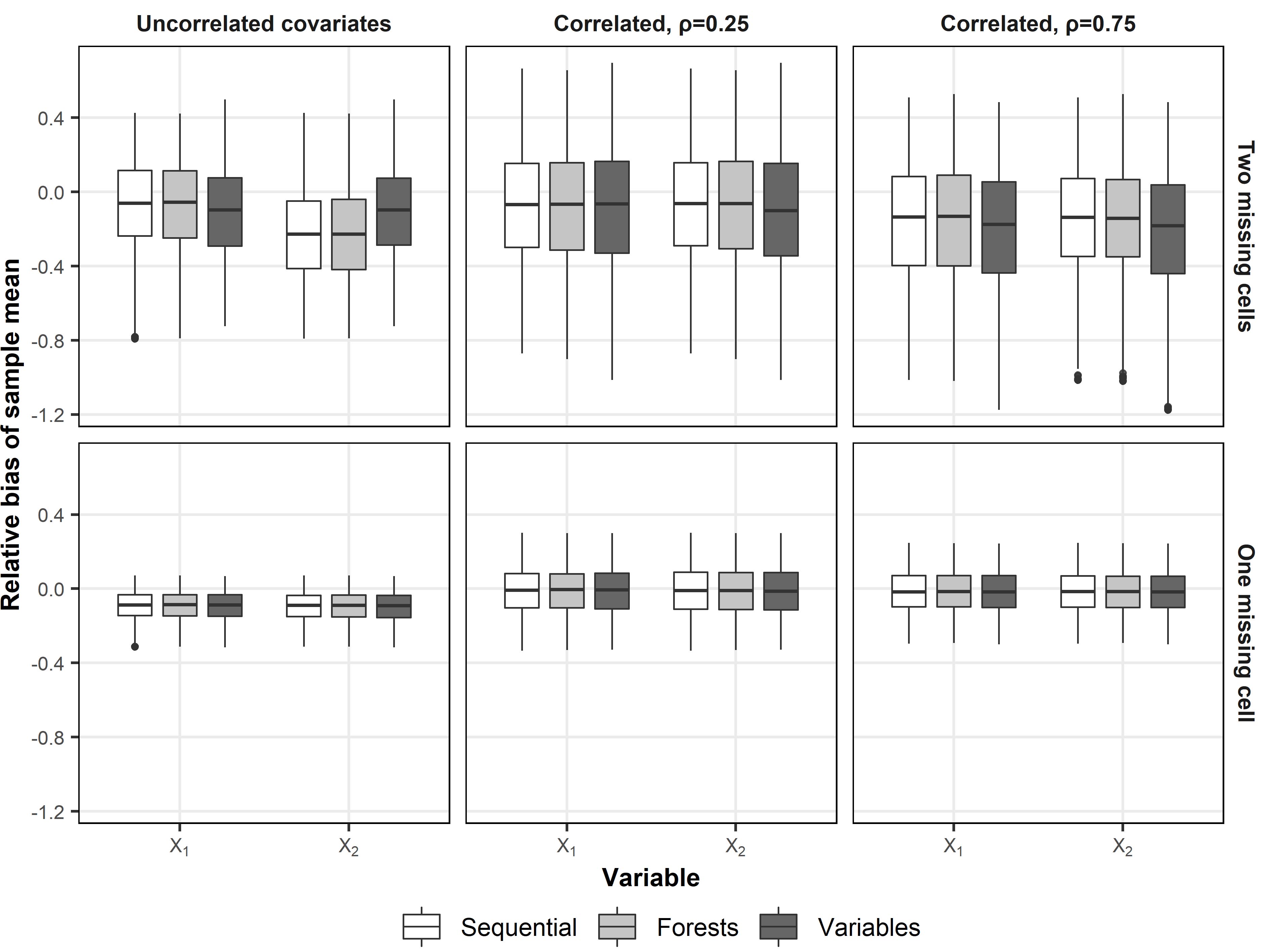}
  \caption{Relative bias of the estimated means of variables}
  \label{fig:2}
\end{figure}
The “variables” parallel imputation strategy resulted in more biased mean
estimates in datasets with multiple missing cells per observation. With two
missing cells per observation for the “uncorrelated covariates” scenario, the
“variables” strategy had an additional downward relative bias when estimating
the mean of $X_{1}$ (median = -9.8\%) compared with the sequential (median =
-6.1\%, p\textless0.001) and “forests” (median = -5.5\%, p\textless0.001)
strategies, while for $X_{2}$ an additional upward relative bias was introduced
(median = -22.8\%, -22.8\%, -9.7\% for “sequential”, “forests”, “variables”,
respectively). For weakly correlated data, the sample mean of $X_{1}$ was
similar (median = -6.9\%, -6.7\%, -6.5\%), but for $X_{2}$ a downward bias was
introduced by “variables” (median = -6.3\%, -6.3\%, -10.0\%). For strongly
correlated data, the “variables” strategy produced biased downward sample
means of $X_{1}$ (median = -13.5\%, -13.2\%, -17.5\%), as well as $X_{2}$
(median = -13.7\%, -14.2\%, -18.2\%). When there was only one missing cell per
observation, the relative bias of the sample mean was similar across the three
strategies (Fig. 2).

\subsection{Relative bias of standard derivation}
\label{sec:3.3}
\begin{figure}[h!]
  \includegraphics[width=\linewidth]{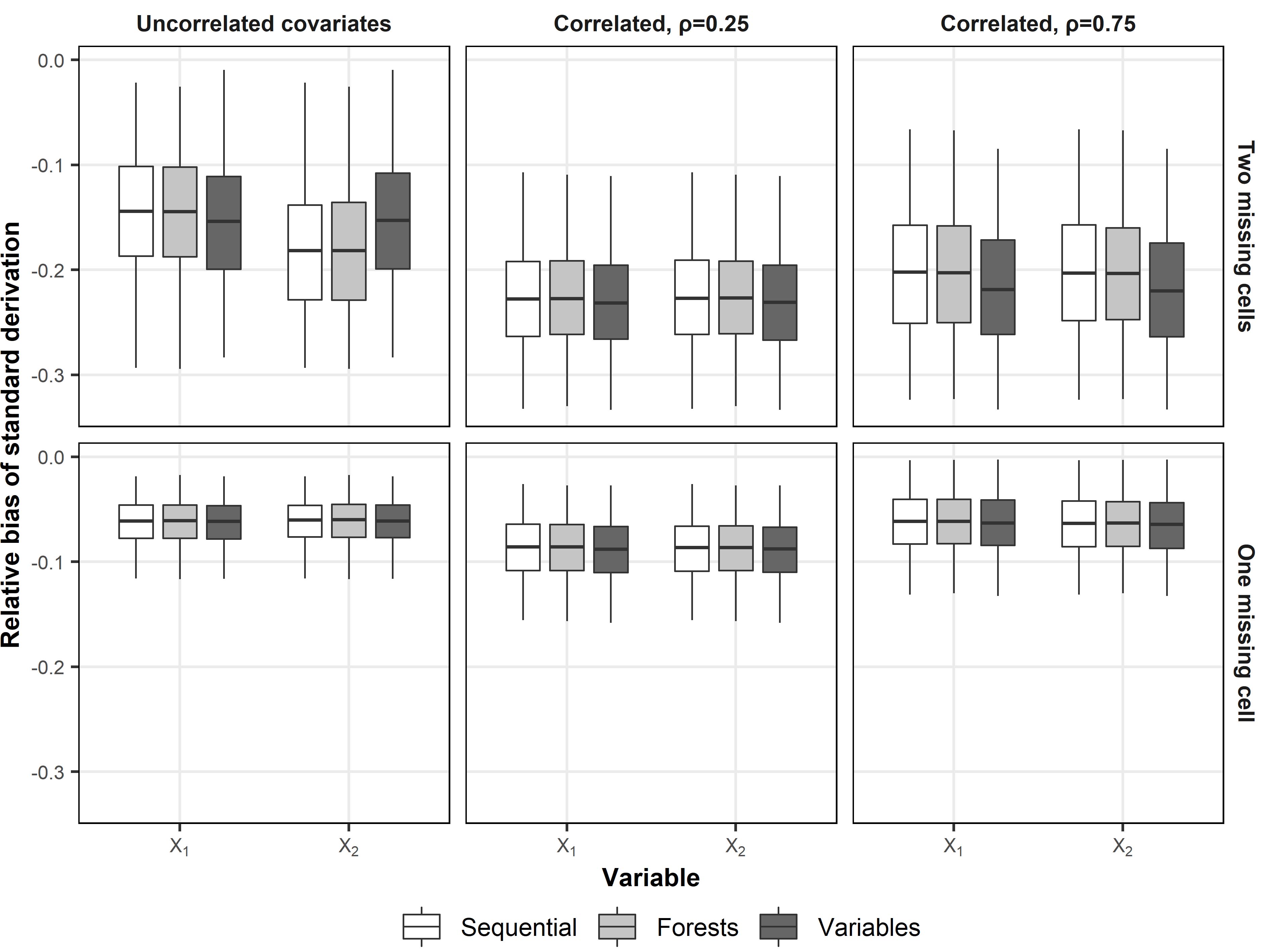}
  \caption{Relative bias of standard derivations of variables}
  \label{fig:3}
\end{figure}
For estimating the standard derivation, all three strategies were systematically
biased downward. The results of the sequential and “forests” strategies were
similar, with the “variables” strategy yielding slightly more biased estimates
(Fig. 3). For example, with two missing cells per observation, the median
relative biases for $X_{1}$ and $X_{2}$ were -14.4\%, -14.4\%, -15.4\% and
-18.2\%, -18.1\%, -15.3\% for “sequential”, “forests”, “variables”,
respectively, for the “uncorrelated covariates” scenario. While for strongly
correlated data, the median relative biases for $X_{1}$ and $X_{2}$ were
-20.2\%, -20.3\%, -21.9\%, and -20.3\%, -20.3\%, -22.0\%, respectively
(Fig. 3).

\subsection{Relative bias of regression coefficient estimates}
\begin{figure}[h!]
  \includegraphics[width=\linewidth]{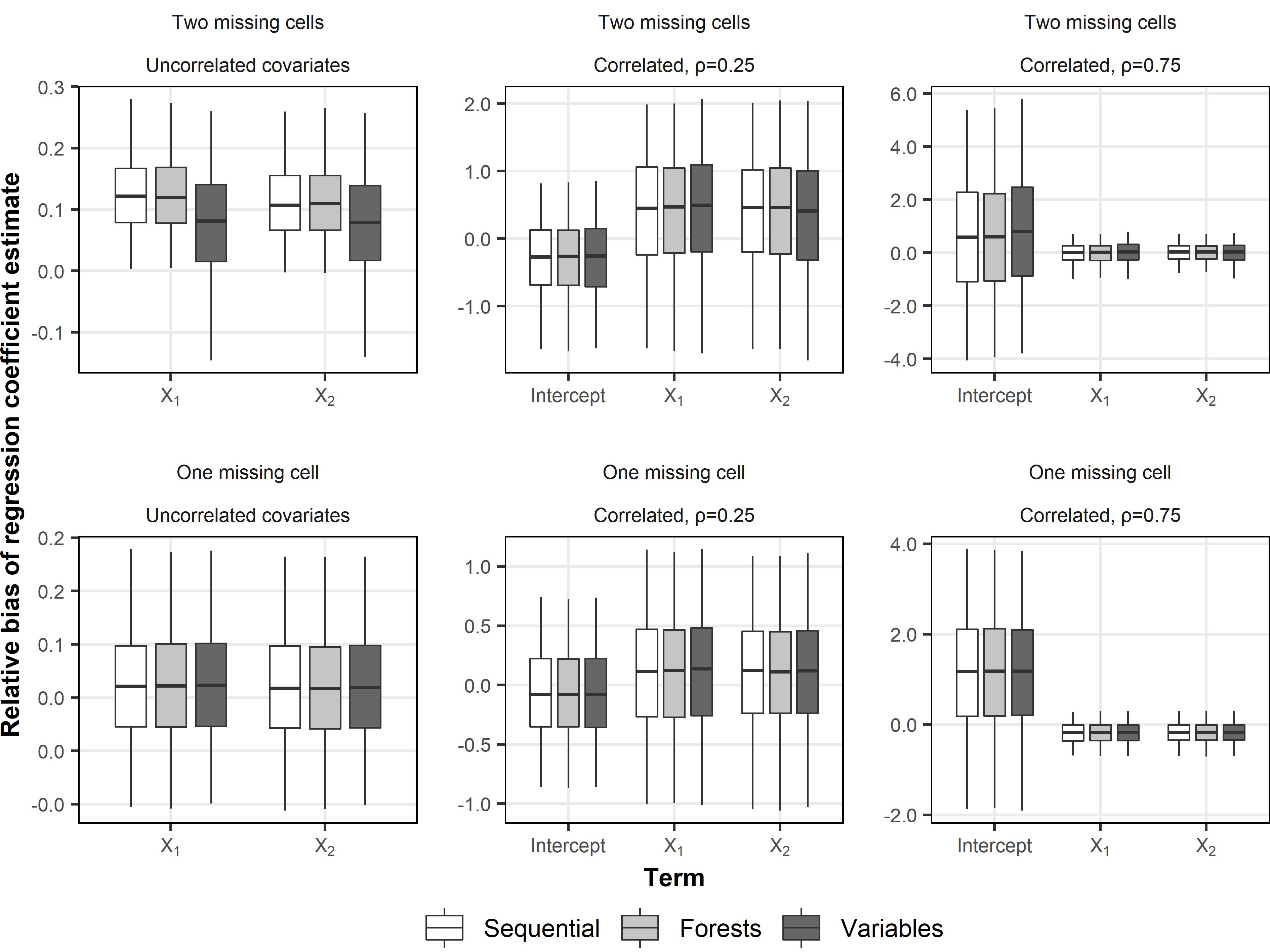}
  \caption{Relative bias of the estimated regression coefficients}
  \label{fig:4}
\end{figure}
MissForest led to biased regression coefficient estimates when covariates are
outcome-dependent MAR, and the “variables” parallel strategy can cause
additional bias. With two missing cells per observation, the “sequential”
and “forests” strategies were similar for the “uncorrelated covariates”
scenario, but the “variables” strategy produced additional downward relative
bias ($X_{1}$: median = 12.1\%, 12.0\%, 8.1\%, for “sequential”, “forests”,
“variables”, respectively; $X_{2}$: median = 10.7\%, 11.0\%, 7.9\%, respectively).
For weakly correlated data, the median relative biases for coefficient estimates
of (intercept, $X_{1}$, $X_{2}$) were (-16.4\%, -15.8\%, -15.5\%),
(9.0\%, 9.4\%, 9.9\%), and (9.1\%, 9.2\%, 8.2\%) for the “sequential”,
“forests”, and “variables” strategies, respectively. While for strongly
correlated data, the median relative biases for coefficient estimates of
(intercept $X_{1}$, $X_{2}$) were (8.3\%, 8.5\%, 11.5\%), (0.2\%, 0.4\%, 1.1\%),
and (1.3\%, 1.2\%, 0.9\%), respectively. For datasets with only one missing cell
per observation, imputation using different parallel strategies gave similar
results (Fig. 4).

\subsection{Bias of correlation between covariates}
\label{sec:3.5}
\begin{figure}[h!]
  \includegraphics[width=\linewidth]{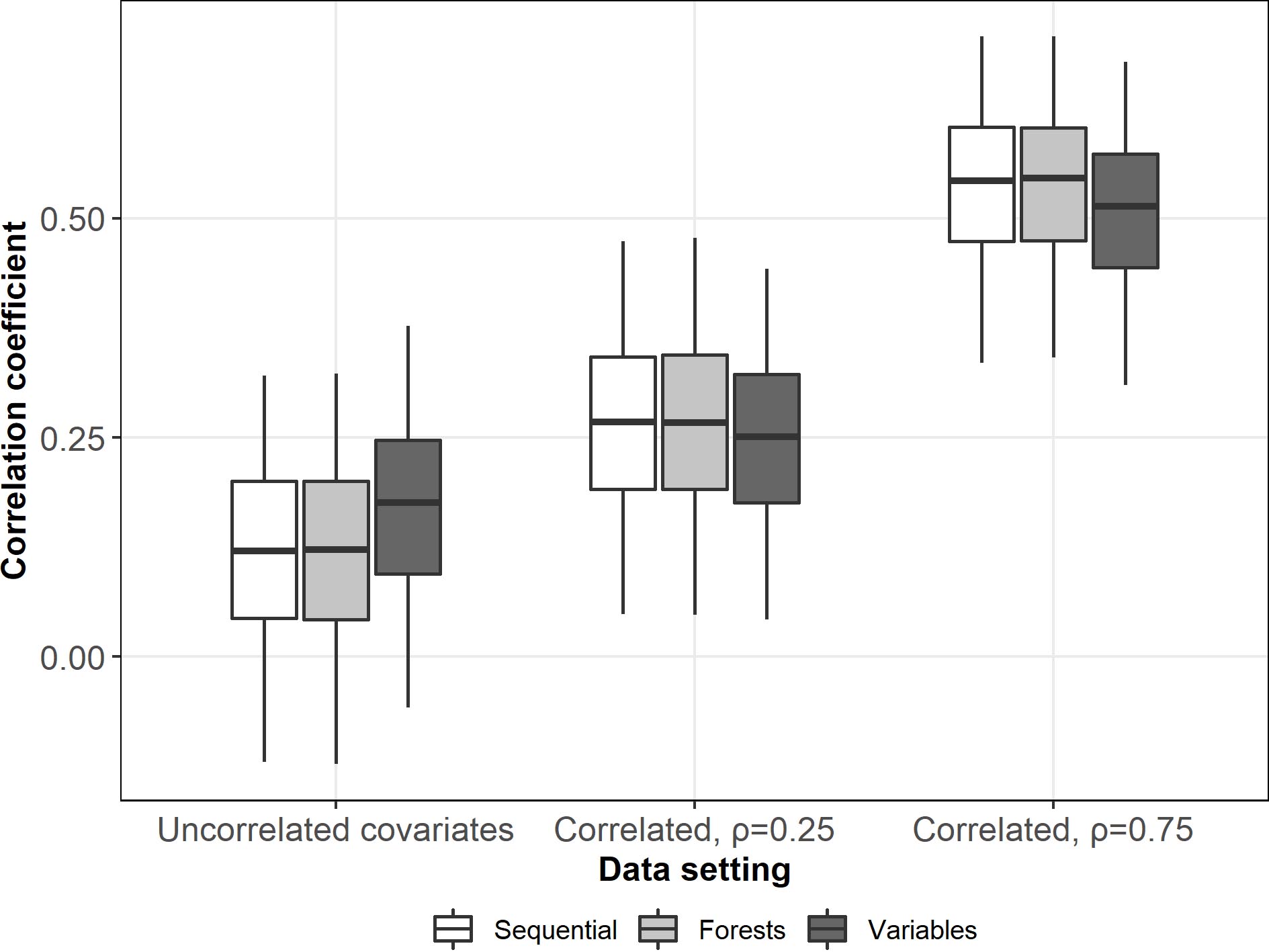}
  \caption{Bias of correlation between covariates}
  \label{fig:5}
\end{figure}
The choice of strategy influenced the correlation between the imputed
covariates. With two missing cells per observation, all three strategies
produced inflated correlations between $X_{1}$ and $X_{2}$ for the
“uncorrelated covariates” scenario (Fig. 5) but the “variables” parallel
strategy resulted in the most biased correlation estimates (median = 0.18
compared to 0.12, for both “sequential” and “forests” strategies). For weakly
correlated data, the correlation coefficients were similar, but for highly
correlated data the correlations were biased downward with the “variables”
strategy yielding the lowest estimate (median = 0.51 compared to 0.55 and 0.54
for “sequential” and “forests”, respectively).

\subsection{Normalized root mean squared error}
\label{sec:3.6}
\begin{figure}[h!]
  \includegraphics[width=\linewidth]{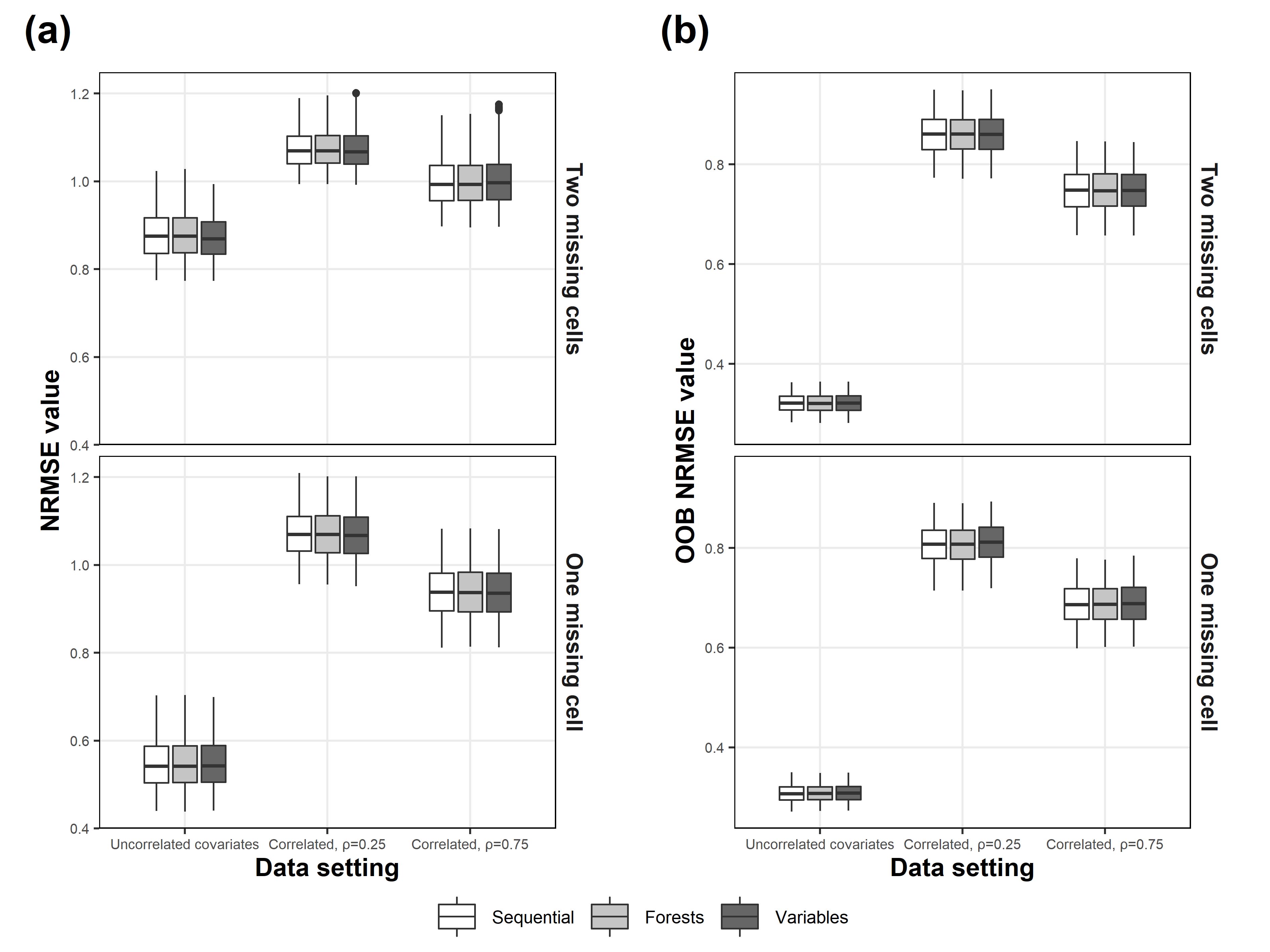}
  \caption{Normalized root mean squared errors (NRMSE) in final imputed data,
  (a) NRMSE calculated based on true data;
  (b) out-of-bag (OOB) errors calculated from OOB observations}
  \label{fig:6}
\end{figure}
The NRMSE values for all three strategies appeared similar, regardless of
whether they were calculated based on the original data (Fig. 6a), or from
the out-of-bag (OOB) values (Fig. 6b).
%
%
%
%
\section{Discussion}
\label{sec:4}
This study examines the parallel algorithms of the RF-based imputation method
missForest and documents for the first time their influence on imputation
results. By distributing imputation computation to multiple computing processes,
reduction in time consumption can be achieved with multi-core processors. In
the “forests” parallel strategy, the imputation for a single variable was
parallelized, and in the “variables” mode, the single iteration of imputation
for different variables was parallelized. Depending on the data structures, our
findings indicate that the “variables” strategy can lead to variations in the
final imputation results compared with the original missForest “sequential”
algorithm. The “variables” strategy yielded additional upward or downward biases
when estimating covariate means, correlation between covariates, and regression
coefficients. This can harm reproducibility and may even lead to false
inference. Moreover, the little variation in NRMSE values between the different
strategies may give a false sense of consistency between them. This also
highlights the fact that evaluating imputation results based solely on NRMSE
values can lead to unreliable conclusions.

The difference in results between the two parallel strategies is a consequence
of their different computing processes. In the parallel “forests” strategy,
the imputation of the current variable is based on the latest state of the
imputation dataset, and the observations of the previously imputed variable
are updated before the start of the imputation of the current variable. The
computation of a single tree in an ensemble is also done independent of other
trees, so this parallel strategy should be similar to the “sequential” strategy
that computes all the trees in an ensemble using a single processor rather than
multiple processors. On the other hand, in the parallel “variables” strategy the
imputation of different variables is done in parallel such that their
computation is based on the same previously updated imputed dataset rather than
variable-wise sequentially updated imputed datasets. This implies that imputed
results are not updated until one cycle of imputation is finished for all the
variables imputed in parallel. Therefore, the imputation of the variables within
a single iteration can no longer be considered sequential, resulting in
different final imputed values from the “sequential” strategy.

The simulations in this study focused on “long” data where the number of
observations is larger than the number of variables in the dataset. However,
for “wide” data with large number of variables, the impact of the non-sequential
updating of imputed values in the “variables” parallel strategy can be even
larger, especially when missing values are scattered across multiple variables
with low inter-correlations. It should be noted that the data settings in this
study were designed to accentuate the differences and consequences of the two
different parallel strategies. In practice, however, datasets like the simulated
data in this study may not be suited for parallel computation as they are not
large enough in terms of number of variables or observations. Also, the
parallelization algorithm can lead to additional time cost,
resulting in more computational time than expected.

This study highlights the importance of thorough testing of computational
algorithms. In particular, it is the lack of technical details in the official
missForest documentations that prompted this investigation. Machine learning
methods like random forests are computationally intensive. Likewise, their
application to big data problems will necessitate the use of parallel
computation algorithms, but developers and users of such statistical software
may be wise to devise simple simulation experiments to test and compare the
algorithms before using them for data analyses. Finally, although we focused
on the missForest method the lessons learned here is not peculiar to it, and
other iterative imputation methods (e.g., MICE) may be faced with similar
problems when adapted for parallel computation.
%
%
%
%
\section{Conclusions}
\label{sec:5}
The problem of using parallel computation has been brought into the forefront
with this study’s investigation of the two parallel strategies implemented in
missForest. It is expected that the proliferation of large datasets and complex
computational methods will continue to fuel the use of parallel algorithms.
The careful analysis of these algorithms is therefore especially important,
and the documentation of these algorithms should include sufficient technical
details and test experiments to inform researchers of potential problems.


\begin{thebibliography}{}
  \bibitem{RefA}
  Liaw A, Wiener M (2002) Classification and regression by randomForest.
  R news 2:18-22
  \bibitem{RefB}
  Mayer M (2019) missRanger: Fast Imputation of Missing Values. CRAN.
  https://CRAN.R-project.org/package=missRanger. Accessed 1 January 2020
  \bibitem{RefC}
  R Core Team (2019) R: A language and environment for statistical computing.
  \bibitem{RefD}
  Ramosaj B, Pauly M (2019) Predicting missing values: a comparative study on
  non-parametric approaches for imputation.
  Computational Statistics 34:1741-1764
  https://doi.org/10.1007/s00180-019-00900-3
  \bibitem{RefE}
  Schouten RM, Lugtig P, Vink G (2018) Generating missing values for simulation
  purposes: a multivariate amputation procedure.
  Journal of Statistical Computation and Simulation 88:2909-2930
  https://doi.org/10.1080/00949655.2018.1491577
  \bibitem{RefF}
  Searle SR, Gruber MHJ (2016) Linear Models. 2nd edn.
  John Wiley \& Sons, Hoboken, New Jersey
  \bibitem{RefG}
  Shah AD, Bartlett JW, Carpenter J, Nicholas O, Hemingway H (2014) Comparison
  of random forest and parametric imputation models for imputing missing data
  using MICE: a CALIBER study.
  American Journal of Epidemiology 179:764-774
  https://doi.org/10.1093/aje/kwt312
  \bibitem{RefH}
  Stekhoven DJ (2013) missForest: Nonparametric Missing Value Imputation using
  Random Forest, R package version 1.4. CRAN.
  https://CRAN.R-project.org/package=missForest. Accessed 1 January 2020
  \bibitem{RefI}
  Stekhoven DJ, Buhlmann P (2012) MissForest--non-parametric missing value
  imputation for mixed-type data.
  Bioinformatics 28:112-118
  https://doi.org/10.1093/bioinformatics/btr597
  \bibitem{RefJ}
  Tang F, Ishwaran H (2017) Random Forest Missing Data Algorithms.
  Statistical Analysis and Data Mining 10:363-377
  https://doi.org/10.1002/sam.11348
  \bibitem{RefK}
  van Buuren S, Groothuis-Oudshoorn K (2011) mice: Multivariate Imputation by
  Chained Equations in R.
  Journal of Statistical Software 45
  https://doi.org/10.18637/jss.v045.i03
  \bibitem{RefL}
  Waljee AK et al. (2013) Comparison of imputation methods for missing
  laboratory data in medicine.
  BMJ Open 3
  https://doi.org/10.1136/bmjopen-2013-002847
  \end{thebibliography}
\end{document}